# Asymmetric pentagonal metal meshes for flexible transparent electrodes and heaters

*Daniel Lordan†, Micheal Burke†, Mary Manning, Andreas Amann, Dan O' Connell, Richard Murphy, Colin Lyons and Aidan J. Quinn\**

Tyndall National Institute, University College Cork, Lee Maltings, Dyke Parade, Cork, Ireland

*Corresponding Author. E-mail: aidan.quinn@tyndall.ie

†Both authors contributed equally to this work.

**ABSTRACT:** Metal meshes have emerged as an important class of flexible transparent electrodes. We report on the characteristics of a new class of asymmetric meshes, tiled using a recently-discovered family of pentagons. Micron-scale meshes were fabricated on flexible polyethylene terephthalate substrates via optical lithography, metal evaporation (Ti 10 nm, Pt 50 nm) and lift-off. Three different designs were assessed, each with the same tessellation pattern and linewidth (5 μm), but with different sizes of the fundamental pentagonal unit. The designs corresponded to areal coverage of the metal patterns of 27% (Design#1), 14% (Design#2) and 9% (Design#3), respectively. Good mechanical stability was observed for both tensile strain and compressive strain. After 1,000 bending cycles, devices subjected to tensile strain showed fractional resistance increases in the range 8% to 17% with the lowest changes observed for Design#2. Devices subjected to compressive strain showed fractional resistance increases in the range 0% to 7% with best results observed for Design#1. The performance of the pentagonal metal mesh devices as visible transparent heaters via Joule heating was also assessed. A saturation temperature of $88 \pm 1$ $^{\circ}$C was achieved at low voltage (5 V) with a fast response time ($\sim 20$ s) and a high thermal resistance ($168 \pm 6$ $^{\circ}$C cm$^2$/W). Finally, de-icing was successfully demonstrated (45 s at 5 V) for an ice layer on a glass coupon placed on top of the PET substrate.

# 1. INTRODUCTION

Transparent conductive electrodes are used in a variety of applications such as thin film solar cells[1], liquid crystal displays,[2] touch panel displays[3] and inorganic/organic light emitting diodes.[4] The current market is dominated by indium-doped tin oxide (ITO) due to its high optical transparency and low sheet resistance.[5] ITO films (700 nm thick) deposited on both rigid and flexible substrates yielded optical transparency between 78% and 85% (averaged over the visible electromagnetic spectrum, not including the substrate contribution), with corresponding sheet resistance, $R_{sheet}$, values between 6 $\Omega$/sq and 9 $\Omega$/sq.[6] However, the brittle nature of ITO, coupled with the rising cost of indium will likely impede its use for future flexible optoelectronic devices.[7-8] These shortcomings have resulted in the investigation of a large number of alternative materials and architectures for flexible transparent electrodes. Potential candidates to date include graphene,[9-10] carbon nanotubes,[11-13] conductive polymers[14-15] and metal nanowire networks.[16-17]

Monolayer graphene has high intrinsic optical transparency of ~ 97.7% (measured transparency values between 97.1-97.5% reported) and good mechanical properties.[10, 18-23] However, monolayer graphene's intrinsic sheet resistance of ~ 6 k $\Omega$/sq is too large for use as a transparent electrode and development of stable adsorbate doping strategies has proved challenging. Graphene's sensitivity to ambient adsorbates, as well as process residue from large area transfer of graphene deposited using chemical vapor deposition (CVD) also present significant barriers to commercial adoption.[24-25] The large range of sheet resistance values reported for high-quality monolayer graphene ($R_{sheet}$ ~ 125-1,200 $\Omega$/sq) reflects these challenges.[10, 20-21]

Carbon nanotubes incorporated into conductive polymer support matrices possess adequate mechanical flexibility and have the potential for low cost fabrication.[26-28] Sheet resistance values in the range of 50-500 $\Omega$/sq have been reported along with transparency values between 63% and 87%, [12, 29-30] depending on nanotube quality (length, diameter and chirality distributions), concentration, doping level, and intra-tube junction resistance, as well as polymer thickness.[4,30-31]

Conductive polymers such as poly(3,4-ethylenedioxythiophene):polystyrenesulfonic acid (PEDOT:PSS) are commercially available in aqueous dispersions and allow cost-effective fabrication by coating or printing methods.[32-33] Sheet resistance values $R_{sheet}$ ~750 $\Omega$/sq and a transparency of 87% have been reported for 100 nm thick PEDOT:PSS films.[34] The conductive properties of these films can be improved by the addition of high boiling point solvents ($R_{sheet}$ ~ 65-176 $\Omega$/sq, T ~ 80-88%)[32] or acids ($R_{sheet}$ ~ 39 $\Omega$/sq, T ~ 80%).[35] Despite the high transparency and mechanical flexibility,[36] polymer films often suffer from unstable sheet resistance due to thermal and environmental stresses.[37]

Metal nanowire networks also allow fabrication of transparent conductive electrodes using solution-based processes. Sheet resistance values between 6.5-38.7 $\Omega$/sq and transparency values between 85-91% have been reported.[16, 38] However, metal nanowire network films have a high surface roughness and high fractional light scatter (i.e. haze) nanowires with large diameters are employed.[39-40]

Mesh-patterned metal films have emerged as promising candidates for the transparent conductive electrode market.[41] The transparency and sheet resistance can be controlled by varying mesh geometry, linewidth, metal thickness as well as employing metals with different resistivity values. The use of linewidths ≤ 5 μm is advantageous due to being undetectable by the naked eye.[42-43] This attribute allows the potential use of metal meshes for applications that require clear visibility.

Recent research efforts on ultra-thin metal meshes have focused on symmetrical geometries such as squares[41, 44] and hexagons[45-46] along with patterning techniques such as UV lithography,[42] nanoimprint lithography[47] and novel techniques such as rolling mask lithography.[48] Ghosh et al. reported square metal meshes (~ 50 nm thick Ni, 20 μm linewidth) on a 2 nm layer of Ni on fused silica substrates patterned by UV lithography with values of $R_{sheet}$ ~ 52 Ω/sq and T ~ 74%.[41] Hexagonal Cu metal meshes (~ 62 nm Cu, 1 μm linewidth) patterned using UV lithography have been reported by Kim et al. with values of $R_{sheet}$ ~ 6.2 Ω/sq and T ~ 91% when an aluminium doped zinc oxide capping layer was applied.[45] Rolith Inc. have fabricated square Al metal meshes (~ 300 nm to 500 nm thick) with sub-micron linewidths (~ 300 nm) using a novel "rolling mask" lithography method, yielding devices with low sheet resistance (~ 3.5 Ω/sq), high transparency (~ 96%) and low haze (4-5%).[49]

While extensive work on symmetrical patterns for metal meshes have been reported, there have been very few reports on asymmetric metal meshes, such as grain boundary lithography. To our knowledge, there have been no reports on uniform asymmetric designs. A new class of asymmetric pentagons which can tile a 2D plane was reported late in 2015.[50] The newly discovered asymmetric pentagon's unit cell consists of a 12 pentagon array and is non-unique. The use of asymmetric metal mesh geometries may have the potential ability to distribute forces when strained leading to improved mechanical stability of these devices. This is particularly important for the integration of flexible optoelectronic devices on deformed or uneven surfaces.

Recently, a promising application of metal meshes as a replacement candidate material for ITO in visible transparent heater technology has been proposed.[5, 51] Visible transparent heaters are used for the de-icing and defrosting of automotive windows, advertisement boards and aviation displays, which require visual transparency in cold environments.[52-55] Besides visual transparency, low sheet resistance is required to achieve a high steady-state "saturation" temperature at acceptable operating voltages.

Here we report on the electrical characteristics of asymmetric metal meshes, tessellated using a new class of pentagons.[50] Three different designs were assessed, each with the same tessellation pattern and linewidth (5 µm), but with different sizes of the fundamental pentagonal unit. Mechanical stability was assessed for both tensile strain and compressive strain. We also report on the performance of the pentagonal metal mesh devices as visible transparent heaters.

## 2. EXPERIMENTAL SECTION

**Fabrication of Pt metal mesh.** The metal mesh devices were patterned on 125 µm thick heat-stabilised polyethylene terephthalate (PET) substrate "Melinex" (Dupont Teijin UK, item # ST504). A 70 mm x 70 mm sized piece was used for processing. Hexamethyldisilazane (HMDS) was spun on the substrate using a Laurell WS400 spinner at 3,000 revolutions per minute (RPM) for 50 s to promote resist adhesion. LOR3A (positive resist) was spun on the substrate at 3,000 RPM for 50 s to produce a nominal thickness of ~ 300 nm followed by baking on a hotplate at 150 $^{o}$C for 3 minutes. Again HMDS was applied at 3,000 RPM for 50 s followed by S1805 at 3,000 RPM for 50 s to produce desired thickness of ~ 450 nm. This was then baked at 115 $^{o}$C for 2 minutes on a hotplate. The substrate was then placed in a Karl Suss MA1006 mask aligner and the wafer was exposed to a dark-field chrome mask (Compugraphics) by ultraviolet (UV) radiation for 3.5 s (exposure dose ~ 35 mJ/cm$^2$). The patterns were developed using MF319 developer for 45 s and immediately placed in deionised water (DI) water to stop the reaction. The substrate was then placed in a Temescal FC2000 electron-beam evaporator system. Prior to evaporation, the chamber was pumped down to ~ 5 x 10$^{-7}$ Torr. A 10 nm Ti adhesion layer was evaporated (at ~ 0.3 nm/s) followed by 50 nm Pt ( ~ 0.5 nm/s). Lift-off of the metal-capped photoresist was achieved by placing the wafer in R1165 Resist Remover at 90 $^{o}$C followed by a DI water rinse and blow drying with nitrogen. Individual mesh devices were of size 7 mm x 11 mm (total die size of 12 mm x 15 mm) with two macro electrodes (7 mm x 2mm) for two-terminal resistance measurements. Four smaller mesh devices of size 2 mm x 2 mm (same linewidth and open area) were utilised for four-terminal sheet resistance measurements.

**Characterisation.** Initial sheet resistance ($R_{sheet}$) values of the Pt mesh devices were evaluated from four-terminal current-voltage measurements performed at room temperature under ambient conditions using an Agilent E5270B parameter analyser interfaced to a LakeShore Desert TTPX probe station (10 mV – 200 mV bias voltage range). The same setup was used for two-terminal resistance measurements. Transparency and fractional light scatter (i.e. haze) data were measured using a UV-vis spectrophotometer (PerkinElmer Lambda 950) over a wavelength range of 400 nm – 800 nm). Quoted transparency values were taken at a wavelength of 550 nm. An integrating sphere setup was utilized to measure the transmitted and scattered light of the mesh devices to evaluate device haze. To test the mechanical stability of the mesh devices, they were manually flexed over a known radius of curvature (~ 3.8 mm) in air. The two-terminal resistance was measured periodically every 200 cycles (up to 1,000 bending cycles). Optical microscopy images were taken using a Leica DMRB microscope in transmission mode at 5x, 10x and 50x magnifications.

To test the viability of the pentagonal mesh device for use as a transparent heater, a thin mist of water was sprayed to a 1 cm x 1cm glass coupon (1.2 mm thick) which was subsequently held above liquid nitrogen vapor. This process was repeated several times until an ice layer thickness ~0.5 mm was observed. The glass substrate was then placed on the mesh device which was connected to a power supply (Aim – TTi EX752M) and a constant bias voltage was then applied. Thermal images and temperature vs time plots were obtained using a FLIR ONE Thermal Imager (120 x 160 pixels resolution, working distance ~ 4.5 cm) interfaced to an Android smartphone. Data analysis was performed using the FLIR Tools software. The accuracy of the FLIR One temperature readings in comparison to a temperature probe (IKA Werke ETS-D4) is given in Table S3 in the supporting information.

## 3. RESULTS AND DISCUSSION

Devices with a fixed linewidth of 5 μm and metal thickness (Ti ~ 10 nm, Pt ~ 50 nm) were fabricated on PET as described in the Experimental Section. The metal mesh devices were based on targeting a lower transmission threshold of 70% for the mesh itself i.e. larger metal area coverage. The transparency of a metal mesh structure, $T_{mesh}$, can be deduced from the geometric design and is approximated as:

$$T_{mesh} \approx 1 - A_{metal}/A_{total} \qquad \qquad \text{Equation 1}$$

where $A_{metal}$ is the area within the unit cell covered by metal and $A_{total}$ is the total unit cell area. A 10x optical microscopy image in transmission mode of the newly discovered asymmetric pentagon design is shown in Figure 1a. For this particular pentagon, the (non-unique) unit cell consists of an array of twelve pentagons (one example shaded in grey). Following equation 1, the expected instrinsic transparency of the pentagonal mesh, $T_{mesh}$, was estimated as :

$$T_{mesh} \approx (1 - 1.36 \ w/d) \qquad \text{(equation 2)},$$

Where $w$ is the linewidth of the mesh and $d$ is the length of the smallest side ($w << d$, see Supporting Information Section S1 for derivation).

The length of side $d$ was varied to assess 3 designs with different transparency values beginning at the lower transmission threshold. The intrinsic transparency values of the individual mesh designs ($T_{mesh}$) were estimated using Equation 2 as $T_{mesh}$ ~ 73% (Design#1, $d$ = 25 μm), $T_{mesh}$ ~

86% (Design#2, $d = 50$ μm) and $T_{mesh} \sim 91\%$ (Design#3, $d = 75$ μm). **Error! Reference source not found.**b shows a photograph taken in natural light of a resultant metal mesh device (Design#3, 9% metal area coverage) patterned on PET. Each die fabricated contains a two-terminal rectangular device (mesh area 7 mm x 11 mm) for resistance and transparency measurements and four small square meshes (2 mm side) with the same mesh parameters as the rectangular device. The sheet resistance (four-terminal) was measured using the smaller mesh devices.

In order to assess the optical properties of the metal mesh devices, the transparency and haze of the metal meshes on PET substrates were measured in the wavelength range of 400 nm to 800 nm as described in the Experimental Section. All absorption spectra were measured vs air. Thus the measured transparency ($T$) includes the absorption by the 125 μm thick PET substrate, $T = T_{PET}$ x $T_{mesh}$. The transparency of bare PET substrates was measured as $\sim 88\%$ at a wavelength, $\lambda$, of 550 nm. The expected transparency of the pentagonal metal mesh devices including the effect of the PET substrate ($T_{PET} \sim 88\%$) were $\sim 64\%$ (design 1, d = 25 μm), $\sim 76\%$ (design 2, d = 50 μm) and $\sim 80\%$ (design 3, d = 75 μm). The measured transparency values of each design (referenced against air) were measured as $\sim 62\%$ (design 1, d = 25 μm) $\sim 74\%$ (design 2, d = 50 μm) and $\sim 78\%$ (design 3, d = 75 μm); see Figure 2. The finite width of the mesh lines and residue from the lithographic process may have contributed to the slightly lower measured values of transparency compared to the estimated values ($\Delta T \sim 2\%$ for all three designs).

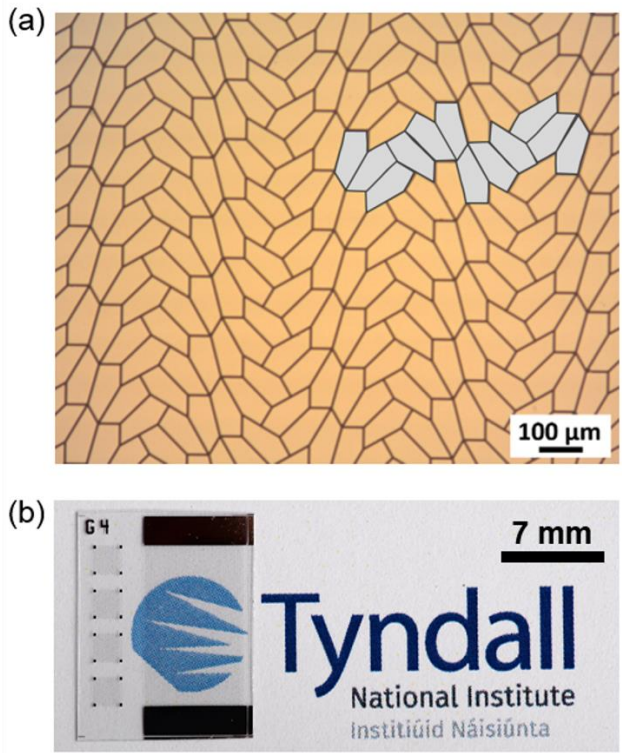

**Figure 1.** (a) High-resolution transmission-mode optical microscopy image showing the asymmetric pentagonal tiling with a linewidth of 5 µm (Design 2, 14% metal area coverage). The gray shaded array of 12 pentagons depicts one (non-unique) unit cell. (b) Photograph taken in daylight of a transparent metal mesh die (Design 3, 9% metal area coverage) on a polyethylene terephthalate (PET) substrate. The 7 mm x 11 mm rectangular mesh with two macro electrodes (7 mm x 2 mm) can be used for two-terminal resistance measurements. Four smaller 2 mm x 2 mm devices were used for four-terminal resistance measurements.

Haze, the ratio of diffusive transmittance (the difference between total and specular transmittance) to total transmittance is another critical parameter for transparent conductive electrodes.[39] For example, a general requirement for touch screen displays are values of haze of < 3%.[54] Larger haze values result in blurriness and reduces clear visibility of the device. The haze of the "Melinex" PET substrates was measured as ~ 1% at λ = 550 nm (Figure 2 inset), in good agreement with the manufacturers' specification of ~ 0.8%.[55] The oscillatory behavior of the transparency and haze spectra in Figure 2 is likely due to optical interference caused by the adhesive coating on one side of the Melinex substrate. The haze of the pentagonal mesh devices, taken at a wavelength of 550 nm were ~ 9% (design 1, 27% metal area coverage), ~ 4% (design 2, 14% metal area coverage), ~ 5% (design 3, 9% metal area coverage). The large haze value of design 1 is expected due to the large metal coverage which can increase the likelihood of light scatter.

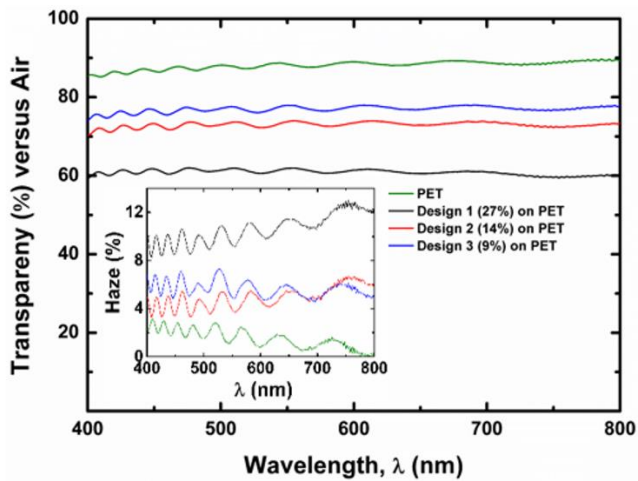

**Figure 2.** Plot of transparency (referenced against air) versus wavelength for all 3 Pt pentagonal metal mesh devices of linewidth 5 μm along with the PET substrate. (Inset) plot of haze versus wavelength for all mesh devices and the PET substrate. The number in brackets represents the % metal area coverage based on the unit cell. The oscillatory nature of transparency and haze spectra is due to the adhesive coating on one side of the PET substrate.

The sheet resistance of mesh patterned metal films is influenced by the mesh geometry, the metal linewidth as well as the thickness and resistivity of the metal used. The 2 mm x 2 mm square mesh devices (**Error! Reference source not found.**b) were used for sheet resistance measurements. The sheet resistance values for the as-fabricated meshes (averaged over seven devices for each design) were ~ 30.3 ± 2.2 Ω/sq (design 1, d = 25 μm), ~ 69.1 ± 4.8 Ω/sq (design 2, d = 50 μm) and ~ 91.7 ± 3.9 Ω/sq. (design 3, d = 75 μm). The two–terminal resistance (averaged over seven devices for a each design) yielded values of ~ 55.7 ± 3.2 Ω (design 1, d = 25 μm), ~ 109.6 ± 7.7 Ω (design 2, d = 50 μm) and ~ 147.5 ± 6.5 Ω (design 3, d = 75 μm).  Two-terminal and four-terminal resistance data for individual devices are provided in Table S1 (Supporting Information).

The ratio of electrical conductivity to optical conductivity is often used as a figure of merit for transparent conductive electrodes,[56-57]

$$\sigma_{DC}/\sigma_{op} = Z_0 / \left( 2R_{sheet} \left( T^{-1/2} - 1 \right) \right)$$
(equation 3),

where $Z_0$ is the impedance of free space (~ 377 Ω) and $R_{sheet}$ and $T$ are the sheet resistance and intrinsic transparency of the transparent electrode material. For the pentagonal meshes using ultrathin Pt (~50 nm), the values for $\sigma_{DC}/\sigma_{op}$ extracted using Equation 3 are 37 ± 3 (design 1, d = 25 μm), 35 ± 2 (design 2, d = 50 μm) and 43 ± 2 (design 3, d = 75 μm). These values are lower than the figure of merit $\sigma_{DC}/\sigma_{op}$ ~ 296 calculated for thicker (700 nm) ITO films on PET (based on $R_{sheet}$ ~ 7 Ω/sq and T ~ 84%).[6] However, the figure of merit for the metal meshes can be further improved by reducing $R_{sheet}$, The sheet resistance for metal grids can be expressed as: [41]

$$R_{sheet} = \xi \ \frac{\rho_G}{t_G FF}$$
(equation 4),

where $\xi$ is a correction factor, $\rho_G$ is the resistivity of the metal, $t_G$ is the thickness of the grid and FF is a geometric filling factor based on the geometry of the metal mesh. Thus, the figure of merit could be improved to match or even exceed the ITO values by using a comparable metal thickness (~ 700 nm) and/or employing a metal with lower resistivity (e.g., Cu).

Mechanical stability is another important criteria for transparent conductive electrodes due to recent consumer demands for flexible electronics. ITO's ceramic nature severely limits its flexibility. Chen et al. reported catastrophic device failure for a ~ 110 nm film at tensile strain values < 1.7% on 188 μm thick PET.[7] Kang et al. reported failure onsets for transparent ITO films on PET (~ 130 μm) when used as heaters at strains at ~ 1.2%.[58] Cyclic bending tests are often undertaken to gauge the effect of repeated strain on device performance. ITO also fails in this regard.[59] Mechanical bending tests (Figure 3) were undertaken on all 3 pentagonal mesh designs. 3 devices of each pentagon design (9 devices in total) were used for tensile bending strain, while 3 devices of each pentagon design (9 devices in total) were used for compressive bending strain. All devices were bent (supported ends) at a radius of curvature, $r_C$~ 3.8 mm. Following the approach of Suo et al., the strain was calculated using the following formula:

$$\varepsilon = \frac{\delta_f + \delta_s}{2r_c} \left[ \frac{1 + 2\eta + \chi\eta^2}{1 + \eta + \chi\eta^2} \right] \text{ (equation 5)}$$

where $\delta_f$ and $\delta_s$ are the thicknesses of the metal (~ 60 nm in total) and substrate (~ 125 μm) respectively, $r_c$ is the radius of curvature (~ 3.8 mm), $\eta = \delta_f/\delta_s$ and $\chi = Y_{Pt}/Y_{PET}$, where $Y_{Pt}$ and $Y_{PET}$ are the Young's moduli of the film and substrate respectively.[60] A strain of ~ 1.6% was calculated using equation 5 for this work using bulk Pt values. The resistance was measured after every 200 bending cycles up to 1,000 bending cycles.

Figure 3a presents the variation of the two terminal resistance after $n$ cycles, $R_n$, in relation to the original two terminal resistance, $R_0$, as a function of bending cycles for all three designs under tensile strain. Good mechanical stability was observed, however all devices showed measurable increases in resistance. After 1,000 bending cycles, devices subjected to tensile strain showed fractional resistance increases in the range 8% to 17% (ie $1.08 < R_{1000}/R_0 < 1.17$) with the lowest changes observed for Design#2. Our results compare favorably with literature reports (see Table S2 in Supporting Information). Kim et al., reported fractional sheet resistance changes < 8% for hexagonal Cu mesh structures (metal thickness ~ 60 nm, linewidth ~ 1 μm) protected with an aluminum doped zinc oxide capping layer (~ 75 nm thick, $R_{sheet}$ ~ 8 kΩ/sq.) after 1,000 bending cycles at a radius of curvature ~ 2 mm.[45]

Devices subjected to compressive strain (Figure 3b) showed lower fractional resistance increases after 1000 cycles than tensile-strained devices, again in agreement with literature reports (see Table S2). Fractional resistance changes in the range 0% to 7% were observed (ie $1 < R_{1000}/R_0 < 1.07$) with best results observed for Design#1. Very recently, Li and co-workers reported an elegant method for fabrication of thick Cu meshes (~ 1.8 μm) with sub-micron linewidths (~900 nm) embedded in cyclic olefin co-polymers.[61] Those structures showed lower resistance variation for compressive strain than tensile strain under a bending radius of 4 mm.

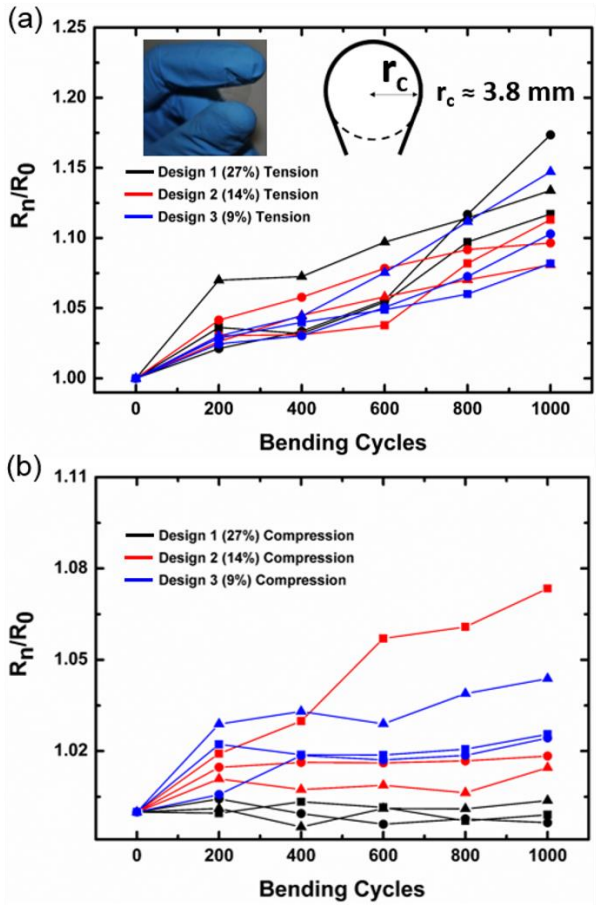

**Figure 3.** (a) Plot of two-terminal resistance variation ($R_n/R_0$) versus number of bending cycles for all 3 pentagonal mesh designs (3 devices of each pentagon design, 9 devices in total) of 5 μm linewidth under tensile strain. (Inset) photo showing bending of the metal mesh on a PET substrate. (b) Plot of two-terminal resistance variation ($R_n/R_0$) versus number of bending cycles for all 3 pentagonal mesh designs of 5 μm linewidth under compressive strain. In both plots, the number in brackets represents the % metal area coverage.

All 3 designs were studied to demonstrate their use as visible transparent heaters. A comparison of the pentagonal mesh devices compared to literature transparent heaters is shown in Table 1.

Although the measured transparency of Design 1 (T ~ 62%) is less than the industrial standard required for automobile windscreens of 70%,[65] its efficacy as a transparent heater was evaluated. A DC bias was applied to the pentagonal mesh devices which caused Joule heating. Constant bias voltages in the range of 2 V to 5 V were applied to the pentagonal mesh devices for 90 s until a saturation temperature range was observed. The exception is design 1 (27% metal area coverage) as voltages above 4 V resulted in temperatures above 120 °C. The FLIR One thermal camera is specified to accurately read temperatures to 120 °C.[66] The temperature was recorded every second using an infrared thermal camera at a fixed working distance of ~ 4.5 cm. A thermal image of the pentagonal mesh (design 2, 14% metal area coverage) depicted in Figure 4a resulted in a max temperature of ~ 86 °C achieved after 90 s seconds by applying 5 V to the device.

For this application, the power dissipated in the device, P, is given as $P = V^2/R$ (equation 6), which implies low device resistance is required to obtain higher temperatures at lower applied voltages. A plot of the temperature versus time (center of the sample) is shown in Figure 4c for design 2 (14% metal area coverage) for each applied voltage (2 V to 5 V) with subsequent saturation temperature ranges of ~ 36 ± 1  °C (2 V), ~ 42 ± 1  °C (2.5 V), ~ 50 ± 1  x °C (3 V), ~ 60 ± 1 °C (3.5 V),  ~ 67 ± 1 °C (4 V), ~ 77 ± 1 °C (4.5 V) and ~ 88 ± 1 °C (5 V). Kang et al. reported two graphene based heater on PET which achieved saturation temperatures of ~ 65 °C (4 layer graphene doped with $HNO_3$) and ~ 100 °C (4 layer graphene doped with $AuCl_3$-$CH_3NO_2$).[58] Although the temperature of ~ 100 °C for the second graphene device is higher than the saturation temperature achieved in this work of ~ 88 ± 1 °C (device 2, 14% metal area coverage), it required an applied voltage of 12V.

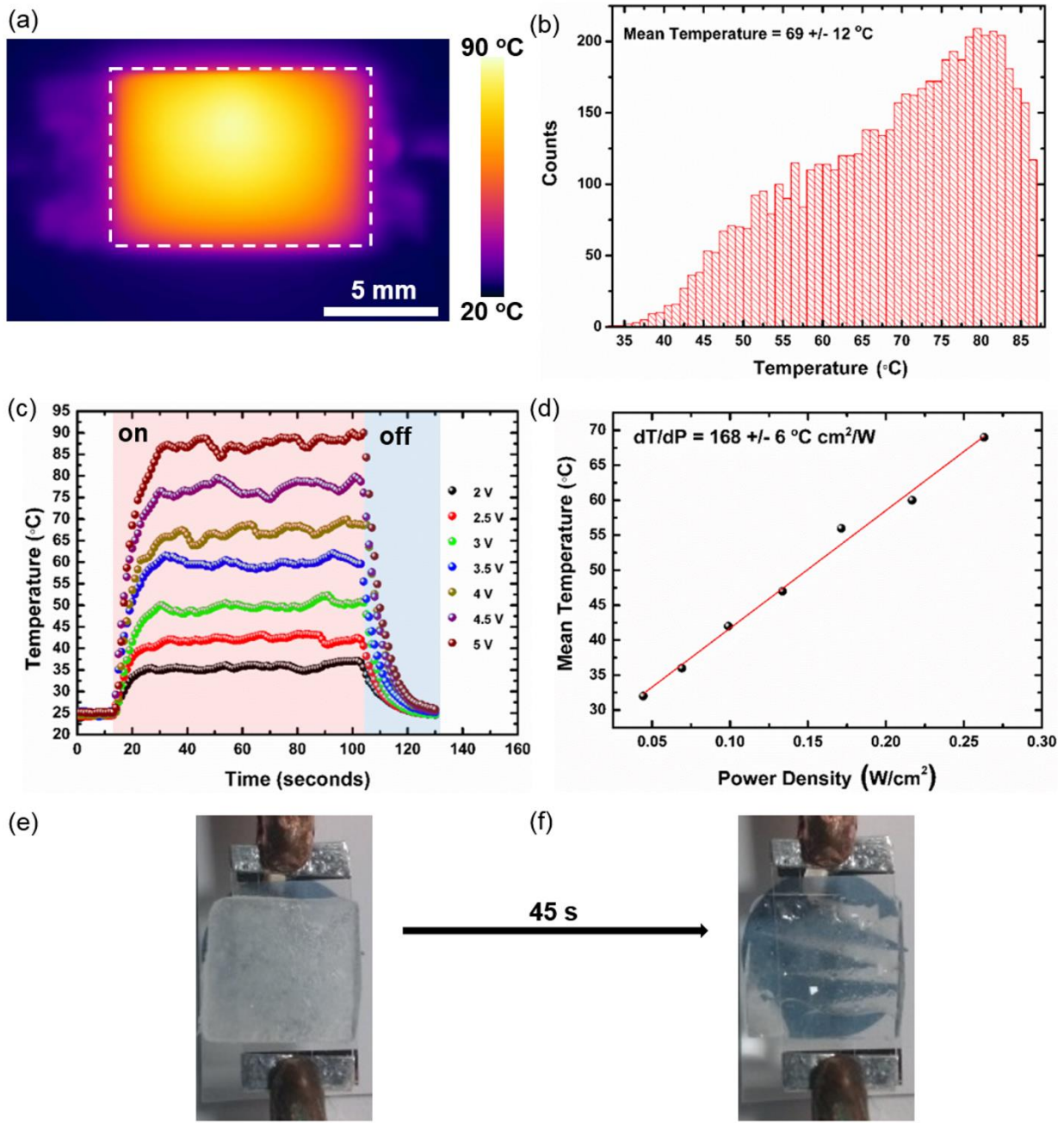

**Figure 4.** (a) Thermal image of pentagonal metal mesh based transparent heater (design 2, 14% metal area coverage) taken at 90 s after applying 5 V. (b) Temperature distribution of the marked area in (a). (c) Temperature versus time of the metal mesh (design 2, 14% metal area coverage) at various voltages (2V to 5V). (d) Plot of mean temperature versus power density for design 2, resulting in a thermal resistance, dT/dP of ~ 168 ºC cm²/W. (e) Ice formation on a glass substrate

of size ~ 1 cm x 1 cm (thickness of ~ 1.2 mm) by liquid nitrogen vapours which was subsequently placed on top of device 2 (14% metal area coverage) on Melinex. (e) Demonstration of ice removal by applying 5 V for ~ 45 seconds, allowing the visibility of the "Tyndall" logo underneath.

**Table 1. Visible Transparent Heater Comparison and De-Icing Parameters[a, b, c, d]**

| Material | Substrate | $T_{material}$[a] (%) | $T_{total}$[b] (%) | $R_{sheet}$ ($\Omega$/sq) | Area (cm$^2$) | Ice Formation | Voltage (V) | Power Density (W/cm$^2$) | Temp. (ºC) | Response time (s) | De-icing time (s) |
|---|---|---|---|---|---|---|---|---|---|---|---|
| Ag mesh[62] | Glass | 77 | 68 | 1 | 10 x 8 | Liquid N$_2$ | 8.5 | 0.57 | 170 | -[c] | 120 |
| Ag mesh[5] | PET | 86 | 76 | 6 | 3.5 x 2.5 | NA[d] | - | 0.2 | 110 | 20 | NA |
| Ag mesh[5] | Convex lens | 86 | 76 | 6 | - | Liquid N$_2$ | 6 | - | 60 | - | 120 |
| Ag mesh -graphene[63] | PET | 78.2 | 69 | 3.8 | 5 x 5 | NA | 4 | 0.003 | 145 | < 30 | NA |
| Ag NW[64] | PET | 90 | 79 | 10 | 5 x 7.5 | NA | 7 | - | 100 | ~ 60 | NA |
| Ag NW[64] | PET | 90 | 79 | 10 | 5 x 7.5 | NA | 5 | - | 70 | ~ 60 | NA |
| Ag NW[64] | Glass | - | | 50 | 5 x 7.5 | Freezer | 12 | - | - | - | 60 |
| Au mesh[51] | Glass | 87 | 77 | 5.4 | 2.5 x 2.5 | NA | 15 | 2.6 | 600 | ~ 38 | NA |
| Design 1_Heater (27% metal area coverage) | PET (125 µm) | 70 | 62 | 29.05 | 0.7 x 1.1 | NA | 4 | 0.39 | 87[e] | ~ 30 | NA |
| Design 2_Heater (14% metal area coverage) | PET (125 µm) | 84 | 74 | 76.64 | 0.7 x 1.1 | Liquid N$_2$ | 5 | 0.26 | 69[d] | ~ 20 | ~ 45 |
| Design 3_Heater (9% metal area coverage) | PET (125 µm) | 89 | 78 | 94.38 | 0.7 x 1.1 | NA | 5 | 0.14 | 56[d] | ~ 20 | NA |

[a]Transparency of underlying material i.e. transparency of mesh, mesh + graphene or nanowires. [b]Total transparency (including the absorbance of the underlying substrate) calculated by using the value of ~ 88% for the PET substrate measured in this work. [c]-, lack of reported data. [d]NA, not applicable. [e,] mean temperature.

Thermal images of the mesh device, 40 s after switching off the power source and thermal images of design 1 and design 3, 90 s after applying 5 V are shown in Figures S2 in the Supporting Information.

A main objective for visible transparent heaters for use as an automobile windscreen de-icer is to provide adequate de-icing temperatures at low input power, i.e. low voltages. A carbon nanotube based heater also required an applied voltage of 12 V to achieve a saturation temperature of ~ 120 °C as reported by Yoon et al.[67] The plots of temperature versus time for design 1 and design 3 are given in Figure S3 in the Supporting Information.

Various groups have investigated alternative heaters based on metal meshes and metal nanowire networks which allow the use of lower applied voltages to achieve suitable saturation temperatures when compared to carbon-based transparent heaters. Kiruthika et al. demonstrated an Ag mesh (~ 300 nm thick) formed from a novel templating method.[62] Voltages of 3 V, 6 V and 9 V were applied with maximum temperatures of ~ 50 °C, ~ 105 °C and ~ 175 °C achieved respectively. The higher saturation temperature achieved by this Ag mesh compared to this work is due to the mesh's low sheet resistance (~ 1 $\Omega$/sq) and the use of higher voltages. The active area of the pentagonal mesh devices herein are 0.77 cm$^2$ and lower temperatures were observed at the edges of the devices which is evident from the thermal image in Figure 4a and temperature distribution plot in Figure 4b with a mean temperature over the marked area of ~ 69 ± 12 °C. Lower temperatures at the edge of a transparent heater has been suggested to occur due to radiative heat loss.[62, 68] Inhomogeneous temperatures at the device periphery has also been reported by Kiruthika et al. where a temperature distribution of ~ 128 ± 43.5 °C was measured over a device with an applied voltage of 9 V. The device area of 10 x 8 cm$^2$ explains the larger temperature distribution compared to the pentagonal metal mesh devices in this work. To prevent temperature variations, the use of a graphene layer in conjunction with a metal mesh has shown to improve temperature homogeneity of the transparent heater due to graphene's high thermal conductivity.[63] Metal nanowire network based transparent heaters require uniformly interconnected nanowires to obtain homogenous temperatures. To achieve uniform interconnects, Kim et al. prepared an Ag nanowire device ($R_S \sim 10$ Ω/sq and T $\sim$ 90%) of size 50 mm x 75 mm.[64] By applying 7 V, the Ag nanowire film achieved a maximum temperature of $\sim 105$ °C. A maximum temperature of $\sim 70$ °C was reported by applying 5 V for this Ag nanowire network. A larger maximum temperature of $\sim 88$ °C was observed for the pentagonal metal mesh (design 2, 14% metal area coverage) using the same applied voltage of 5V.

Another parameter for evaluating the performance of a visible transparent heater is the response time. The response time is defined as the time taken to achieve saturation temperature.[54] A low response time is desired at low input power. An automobile battery voltage is commonly 12 V, with voltages less than this value required for this heater application, to minimize power consumption. The response time increased from $\sim 10$ s to $\sim 20$ s when the voltage was increased from 2 V to 5 V for design 2 (14% metal area coverage). For example, a crackle templated Ag mesh had a response time of $\sim 170$ s.[62] But this is likely due to the use a thick underlying glass substrate ($\sim 1.5$ mm thick) and that the temperature was recorded from the backside of the device. The low response time in this work compares favourably with graphene-based[58] (response time $\sim$ 100 s) and nanowire network-based[64] (response time $\sim 60$ s) transparent heaters which required higher voltages of 12 V and 7 V respectively. The graphene-based heaters with a response time of $\sim 100$ s is likely due to the use of an additional PET layer on both graphene devices, which was used to protect the device against atmospheric environmental stresses. After 90 s of applied voltage to the pentagonal mesh devices in this work, the power supply was turned off and the temperature and time were recorded. The devices returned to room temperature in < 40 s for each applied voltage. The thermal resistance, dT/dP, of all 3 pentagonal metal mesh devices were obtained from the slope of the temperature versus power density plots and were found to be $\sim 153$ °C cm$^2$/W

(design 1, 27%), ~ 168 ºC cm$^2$/W (design 2, 14%) and ~ 190 ºC cm$^2$/W (design 3, 9%). A large thermal resistance is desired for transparent heater applications. The state of the art values for thermal resistances in the current literature are ~ 515 ºC cm$^2$/W and ~ 255.2 ºC cm$^2$/W.[5, 62] The thermal resistances quoted herein are based on the mean temperature value achieved over the heater area at 90s of applied voltage (see Figure S4 in the Supporting Information for plots of mean temperature versus power density for design 1 and design 2). Since heat loss is expected at the edges of a transparent heater, the thermal resistance quoted at maximum temperature is not a true reflection of the heater's performance. However, plots of the maximum temperature achieved after 90 s of applied voltage versus power density along with the thermal resistances at maximum temperature for each pentagon design are shown in Figure S5 in the Supporting Information. Design 3 has a higher thermal resistance that design 2 (190 ºC cm$^2$/W versus 168 ºC cm$^2$/W) but its mean temperature at 5 V was only ~ 56 ºC compared to a mean temperature of ~ 69 ºC for design 2 at 5 V. The thermal resistance of design 2 of ~ 168 ºC cm$^2$/W is higher than the values obtained for other heaters reported in the literature based on single walled carbon nanotubes (~ 140 ºC cm$^2$/W),[69] graphene (~ 163 ºC cm$^2$/W)[70] and Ag nanowires (~ 85 ºC cm$^2$/W),[52] and compares well to the thermal resistance achieved for an Au wire network transparent heater (~ 189 ºC cm$^2$/W).[51]

The Joule effect was used to demonstrate the use of the asymmetric pentagonal metal mesh as a visible transparent heater for de-icing purposes. For its use of de-icing windscreens, the mesh device would be embedded in the glass rather than being fabricated on the glass surface itself where delamination can occur. Therefore using a frozen glass piece which is placed on the mesh device followed by application of a voltage is a better reflection of the transparent heater's de-icing performance. Many publications have carried out de-icing on films where ice formation is achieved by placing the film in a freezer.[64, 67] In this work, a similar reported approach was used,[5, 62], which consisted of applying a thin layer of water over liquid nitrogen vapours with sequential addition of a thin water layer using a spray bottle. Before the voltage was applied to the mesh device, the temperature ($\sim$ 19.6 °C) and humidity ($\sim$ 56 %) in the lab were recorded. Once the glass piece with frozen water was placed on the mesh device (design 2, 14% metal area coverage), a DC bias of 5 V was applied. The de-icing process took $\sim$ 45 s and the visualisation of the "Tyndall" logo after de-icing is evident (Figure 4f). The time taken to de-ice compares favourably with the similar methods undertaken by Kiruthika et al. and Gupta et al. where de-icing occurred in $\sim$ 120 s under 8.5 V and $\sim$ 120 s under 6 V respectively. Larger voltages and longer de-icing times were required as the voltage was applied in the presence of liquid nitrogen vapours.

## 4. Conclusion

We have demonstrated transparent platinum mesh electrodes and heaters based on a newly discovered asymmetric pentagonal tiling. Device performance compares well with literature reports for flexible transparent electrodes and heaters. Future work will focus on investigating the influence of asymmetry on failure under multi-axial strain and design optimization targeting highly reliable or even self-healing flexible, transparent electrodes.

ASSOCIATED CONTENT

**Supporting Information**

A derivation of the transparency for the pentagonal metal mesh, a table of initial two terminal and four terminal resistance and two terminal resistance after each 200 bending cycles, a literature table of cyclic bending for metal meshes, accuracy of FLIR One temperature readings, thermal image of design 2 40 s after removing the voltage, thermal images of designs 1 and 3 at 90 s after applying 4 V (design 1) and 5 V (design 3), temperature versus time for design 1 (voltages in the range of 2 V to 4 V in steps of 0.5 V) and design 3 (voltages in the range of 2 V to 5 V in steps of 0.5 V), mean temperature versus power density for design 1 and design 3 and maximum temperature versus power density for all 3 pentagon designs. This material is available free of charge via the Internet at http://pubs.acs.org.

AUTHOR INFORMATION

**Corresponding Author**

*E-mail: aidan.quinn@tyndall.ie.

**Author Contributions**

[†]Both authors contributed equally to this work.

**Notes**

The author's declare no competing financial interest.

ACKNOWLEDGMENT

This work was supported by the EU under the FP7 project "GONEXTs" (309201). The author would like to thank Thane C. Gough, business development manager of Flexible Electronics Dupont Teijin Films (UK) Ltd who supplied free samples of the flexible PET substrate "Melinex" used in this work.

**REFERENCES**

1. Rowell, M. W.; McGehee, M. D. Transparent Electrode Requirements for Thin Film Solar Cell Modules. Energy & Environmental Science 2011, 4, 131-134.

2. Ellmer, K. Past Achievements and Future Challenges in the Development of Optically Transparent Electrodes. *Nature Photonics* **2012,** *6*, 809-817.

3. Madaria, A. R.; Kumar, A.; Zhou, C. Large Scale, Highly Conductive and Patterned Transparent Films of Silver Nanowires on Arbitrary Substrates and Their Application in Touch Screens. *Nanotechnology* **2011,** *22*, 245201.

4. Hecht, D. S.; Hu, L.; Irvin, G. Emerging Transparent Electrodes Based on Thin Films of Carbon Nanotubes, Graphene, and Metallic Nanostructures. *Adv. Mater.* **2011,** *23*, 1482-1513.

5. Gupta, R.; Rao, K. D. M.; Srivastava, K.; Kumar, A.; Kiruthika, S.; Kulkarni, G. U. Spray Coating of Crack Templates for the Fabrication of Transparent Conductors and Heaters on Flat and Curved Surfaces. *ACS Applied Materials & Interfaces* **2014,** *6*, 13688-13696.

6. Guillén, C.; Herrero, J. Comparison Study of Ito Thin Films Deposited by Sputtering at Room Temperature onto Polymer and Glass Substrates. *Thin Solid Films* **2005,** *480-481*, 129-132.

7. Chen, Z.; Cotterell, B.; Wang, W.; Guenther, E.; Chua, S. J. A Mechanical Assessment of Flexible Optoelectronic Devices. *Thin Solid Films* **2001,** *394*, 202-206.

8. Leterrier, Y.; Médico, L.; Demarco, F.; Månson, J. A. E.; Betz, U.; Escolà, M. F.; Olsson, M. K.; Atamny, F. Mechanical Integrity of Transparent Conductive Oxide Films for Flexible Polymer-Based Displays. *Thin Solid Films* **2004,** *460*, 156-166.

9.      Kim, D.; Lee, D.; Lee, Y.; Jeon, D. Y. Work-Function Engineering of Graphene Anode by Bis(Trifluoromethanesulfonyl)Amide Doping for Efficient Polymer Light-Emitting Diodes. *Adv. Funct. Mater.* **2013**, *23*, 5049-5055.

10.     Bae, S.; Kim, H.; Lee, Y.; Xu, X.; Park, J. S.; Zheng, Y.; Balakrishnan, J.; Lei, T.; Ri Kim, H.; Song, Y. I.; Kim, Y. J.; Kim, K. S.; Özyilmaz, B.; Ahn, J. H.; Hong, B. H.; Iijima, S. Roll-to-Roll Production of 30-Inch Graphene Films for Transparent Electrodes. *Nature Nanotechnology* **2010,** *5*, 574-578.

11.     Dan, B.; Irvin, G. C.; Pasquali, M. Continuous and Scalable Fabrication of Transparent Conducting Carbon Nanotube Films. *ACS Nano* **2009,** *3*, 835-843.

12.     Zhang, D.; Ryu, K.; Liu, X.; Polikarpov, E.; Ly, J.; Tompson, M. E.; Zhou, C. Transparent, Conductive, and Flexible Carbon Nanotube Films and Their Application in Organic Light-Emitting Diodes. *Nano Lett.* **2006,** *6*, 1880-1886.

13.     Hecht, D. S.; Heintz, A. M.; Lee, R.; Hu, L.; Moore, B.; Cucksey, C.; Risser, S. High Conductivity Transparent Carbon Nanotube Films Deposited from Superacid. *Nanotechnology* **2011,** *22*, 075201.

14.     Jönsson, S. K. M.; Birgerson, J.; Crispin, X.; Greczynski, G.; Osikowicz, W.; Denier van der Gon, A. W.; Salaneck, W. R.; Fahlman, M. The Effects of Solvents on the Morphology and Sheet Resistance in Poly(3,4-Ethylenedioxythiophene)–Polystyrenesulfonic Acid (Pedot–Pss) Films. *Synth. Met.* **2003,** *139*, 1-10.

15.     Kim, N.; Kee, S.; Lee, S. H.; Lee, B. H.; Kahng, Y. H.; Jo, Y.-R.; Kim, B.-J.; Lee, K. Highly Conductive Pedot:Pss Nanofibrils Induced by Solution-Processed Crystallization. *Adv. Mater.* **2014,** *26*, 2268-2272.

16.     De, S.; Higgins, T. M.; Lyons, P. E.; Doherty, E. M.; Nirmalraj, P. N.; Blau, W. J.; Boland, J. J.; Coleman, J. N. Silver Nanowire Networks as Flexible, Transparent, Conducting Films: Extremely High Dc to Optical Conductivity Ratios. *ACS Nano* **2009,** *3*, 1767-1774.

17.     Liu, C.-H.; Yu, X. Silver Nanowire-Based Transparent, Flexible, and Conductive Thin Film. *Nanoscale Research Letters* **2011,** *6*, 1-8.

18.     Nair, R. R.; Blake, P.; Grigorenko, A. N.; Novoselov, K. S.; Booth, T. J.; Stauber, T.; Peres, N. M. R.; Geim, A. K. Fine Structure Constant Defines Visual Transparency of Graphene. *Science* **2008,** *320*, 1308-1308.

19.     Bonaccorso, F.; Sun, Z.; Hasan, T.; Ferrari, A. C. Graphene Photonics and Optoelectronics. *Nature Photonics* **2010,** *4*, 611-622.

20.     Suk, J. W.; Kitt, A.; Magnuson, C. W.; Hao, Y.; Ahmed, S.; An, J.; Swan, A. K.; Goldberg, B. B.; Ruoff, R. S. Transfer of Cvd-Grown Monolayer Graphene onto Arbitrary Substrates. *ACS Nano* **2011,** *5*, 6916-6924.

21.     Sun, Z.; Yan, Z.; Yao, J.; Beitler, E.; Zhu, Y.; Tour, J. M. Growth of Graphene from Solid Carbon Sources. *Nature* **2010,** *468*, 549-552.

22.     Lee, C.; Wei, X.; Kysar, J. W.; Hone, J. Measurement of the Elastic Properties and Intrinsic Strength of Monolayer Graphene. *Science* **2008,** *321*, 385-388.

23.     Kim, K. S.; Zhao, Y.; Jang, H.; Lee, S. Y.; Kim, J. M.; Kim, K. S.; Ahn, J. H.; Kim, P.; Choi, J. Y.; Hong, B. H. Large-Scale Pattern Growth of Graphene Films for Stretchable Transparent Electrodes. *Nature* **2009,** *457*, 706-710.

24.     Novoselov, K. S.; Geim, A. K.; Morozov, S. V.; Jiang, D.; Katsnelson, M. I.; Grigorieva, I. V.; Dubonos, S. V.; Firsov, A. A. Two-Dimensional Gas of Massless Dirac Fermions in Graphene. *Nature* **2005,** *438*, 197-200.

25.    Chan, J.; Venugopal, A.; Pirkle, A.; McDonnell, S.; Hinojos, D.; Magnuson, C. W.; Ruoff, R. S.; Colombo, L.; Wallace, R. M.; Vogel, E. M. Reducing Extrinsic Performance-Limiting Factors in Graphene Grown by Chemical Vapor Deposition. *ACS Nano* **2012,** *6,* 3224-3229.

26.    Snow, E. S.; Campbell, P. M.; Ancona, M. G.; Novak, J. P. High-Mobility Carbon-Nanotube Thin-Film Transistors on a Polymeric Substrate. *Appl. Phys. Lett.* **2005,** *86,* 1-3.

27.    Cao, Q.; Rogers, J. A. Ultrathin Films of Single-Walled Carbon Nanotubes for Electronics and Sensors: A Review of Fundamental and Applied Aspects. *Adv. Mater.* **2009,** *21,* 29-53.

28.    Rouhi, N.; Jain, D.; Burke, P. J. High-Performance Semiconducting Nanotube Inks: Progress and Prospects. *ACS Nano* **2011,** *5,* 8471-8487.

29.    Yu, Z.; Niu, X.; Liu, Z.; Pei, Q. Intrinsically Stretchable Polymer Light-Emitting Devices Using Carbon Nanotube-Polymer Composite Electrodes. *Adv. Mater.* **2011,** *23,* 3989-3994.

30.    Rowell, M. W.; Topinka, M. A.; McGehee, M. D.; Prall, H.-J.; Dennler, G.; Sariciftci, N. S.; Hu, L.; Gruner, G. Organic Solar Cells with Carbon Nanotube Network Electrodes. *Appl. Phys. Lett.* **2006,** *88,* 233506.

31.    Yao, Z.; Postma, H. W. C.; Balents, L.; Dekker, C. Carbon Nanotube Intramolecular Junctions. *Nature* **1999,** *402,* 273-276.

32.    Kim, Y. H.; Sachse, C.; MacHala, M. L.; May, C.; Müller-Meskamp, L.; Leo, K. Highly Conductive Pedot:Pss Electrode with Optimized Solvent and Thermal Post-Treatment for Ito-Free Organic Solar Cells. *Adv. Funct. Mater.* **2011,** *21,* 1076-1081.

33.    Xiong, Z.; Liu, C. Optimization of Inkjet Printed Pedot:Pss Thin Films through Annealing Processes. *Organic Electronics: physics, materials, applications* **2012,** *13,* 1532-1540.

34.     Lipomi, D. J.; Lee, J. A.; Vosgueritchian, M.; Tee, B. C. K.; Bolander, J. A.; Bao, Z. Electronic Properties of Transparent Conductive Films of Pedot:Pss on Stretchable Substrates. *Chem. Mater.* **2012,** *24*, 373-382.

35.     Xia, Y.; Sun, K.; Ouyang, J. Solution-Processed Metallic Conducting Polymer Films as Transparent Electrode of Optoelectronic Devices. *Adv. Mater.* **2012,** *24*, 2436-2440.

36.     Cho, C. K.; Hwang, W. J.; Eun, K.; Choa, S. H.; Na, S. I.; Kim, H. K. Mechanical Flexibility of Transparent Pedot:Pss Electrodes Prepared by Gravure Printing for Flexible Organic Solar Cells. *Sol. Energy Mater. Sol. Cells* **2011,** *95*, 3269-3275.

37.     Vitoratos, E.; Sakkopoulos, S.; Dalas, E.; Paliatsas, N.; Karageorgopoulos, D.; Petraki, F.; Kennou, S.; Choulis, S. A. Thermal Degradation Mechanisms of Pedot:Pss. *Org. Electron.* **2009,** *10*, 61-66.

38.     van de Groep, J.; Spinelli, P.; Polman, A. Transparent Conducting Silver Nanowire Networks. *Nano Lett.* **2012,** *12*, 3138-3144.

39.     Preston, C.; Xu, Y.; Han, X.; Munday, J. N.; Hu, L. Optical Haze of Transparent and Conductive Silver Nanowire Films. *Nano Research* **2013,** *6*, 461-468.

40.     Hu, L.; Kim, H. S.; Lee, J.-Y.; Peumans, P.; Cui, Y. Scalable Coating and Properties of Transparent, Flexible, Silver Nanowire Electrodes. *ACS Nano* **2010,** *4*, 2955-2963.

41.     Ghosh, D. S.; Chen, T. L.; Pruneri, V. High Figure-of-Merit Ultrathin Metal Transparent Electrodes Incorporating a Conductive Grid. *Appl. Phys. Lett.* **2010,** *96*, 041109.

42.     Sam, F. L. M.; Mills, C. A.; Rozanski, L. J.; Silva, S. R. P. Thin Film Hexagonal Gold Grids as Transparent Conducting Electrodes in Organic Light Emitting Diodes. *Laser and Photonics Reviews* **2014,** *8*, 172-179.

43.     Philipp, H. Touch Screen Sensor. Patent US20100123670 A1, May 20, 2010.

44.    Dong, P.; Zhu, Y.; Zhang, J.; Peng, C.; Yan, Z.; Li, L.; Peng, Z.; Ruan, G.; Xiao, W.; Lin, H.; Tour, J. M.; Lou, J. Graphene on Metal Grids as the Transparent Conductive Material for Dye Sensitized Solar Cell. *Journal of Physical Chemistry C* **2014,** *118*, 25863-25868.

45.    Kim, W. K.; Lee, S.; Hee Lee, D.; Hee Park, I.; Seong Bae, J.; Woo Lee, T.; Kim, J. Y.; Hun Park, J.; Chan Cho, Y.; Ryong Cho, C.; Jeong, S. Y. Cu Mesh for Flexible Transparent Conductive Electrodes. *Sci. Rep.* **2015,** *5*, 3519−3521.

46.    Galagan, Y.; J.M. Rubingh, J.-E.; Andriessen, R.; Fan, C.-C.; W.M. Blom, P.; C. Veenstra, S.; M. Kroon, J. Ito-Free Flexible Organic Solar Cells with Printed Current Collecting Grids. *Sol. Energy Mater. Sol. Cells* **2011,** *95*, 1339-1343.

47.    Kang, M. G.; Guo, L. J. Nanoimprinted Semitransparent Metal Electrodes and Their Application in Organic Light-Emitting Diodes. *Adv. Mater.* **2007,** *19*, 1391-1396.

48.    Seitz, O.; Geddes Iii, J. B.; Aryal, M.; Perez, J.; Wassei, J.; McMackin, I.; Kobrin, B. In *Antireflective Surface Patterned by Rolling Mask Lithography*, Proceedings of SPIE - The International Society for Optical Engineering, **2014**.

49.    Aryal, M.; Geddes, J.; Seitz, O.; Wassei, J.; McMackin, I.; Kobrin, B. Sub-Micron Transparent Metal Mesh Conductor for Touch Screen Displays. *Digest of Technical Papers - SID International Symposium* **2014,** *45*, 194-196.

50.    Sugimito, T., Tiling Problem: Convex Pentagons for Edge-to-Edge Monohedral Tiling and Convex Polygons for Aperiodic Tiling. 2015, arXiv:metric geometry/ 1508.01864v3 arXiv.org e-Print archive. https://arxiv.org/abs/1508.01864v3 (accessed May 16, 2016)

51.    Rao, K. D. M.; Kulkarni, G. U. A Highly Crystalline Single Au Wire Network as a High Temperature Transparent Heater. *Nanoscale* **2014,** *6*, 5645-5651.

52.      Celle, C.; Mayousse, C.; Moreau, E.; Basti, H.; Carella, A.; Simonato, J.-P. Highly Flexible Transparent Film Heaters Based on Random Networks of Silver Nanowires. *Nano Research* **2012,** *5*, 427-433.

53.      Peterson, J. D. Control System for Eliminating Ice from a Transparent Windshield Panel. Patent US2806118 A, Sep 10, 1957.

54.      Gupta, R.; Rao, K. D. M.; Kiruthika, S.; Kulkarni, G. U. Visibly Transparent Heaters. *ACS Applied Materials and Interfaces* **2016,** *8*, 12559-12575.

55.      Melinex ; Product No. St504 [Online]; Flexible Electronics Dupont Teijin Films; USA: Chester,                                                                                   Va. Http://Www.Dupontteijinfilms.Com/Filmenterprise/Datasheet.Asp?Result=Print&Id=269&Versi on=Us (Accessed May 10, 2016).

56.      Sepulveda-Mora, S. B.; Cloutier, S. G. Figures of Merit for High-Performance Transparent Electrodes Using Dip-Coated Silver Nanowire Networks. *Journal of Nanomaterials* **2012,** *2012*, 7.

57.      De, S.; Coleman, J. N. Are There Fundamental Limitations on the Sheet Resistance and Transmittance of Thin Graphene Films? *ACS Nano* **2010,** *4*, 2713-2720.

58.      Kang, J.; Kim, H.; Kim, K. S.; Lee, S.-K.; Bae, S.; Ahn, J.-H.; Kim, Y.-J.; Choi, J.-B.; Hong, B. H. High-Performance Graphene-Based Transparent Flexible Heaters. *Nano Lett.* **2011,** *11*, 5154-5158.

59.      Cairns, D. R.; Crawford, G. P. Electromechanical Properties of Transparent Conducting Substrates for Flexible Electronic Displays. *Proc. IEEE* **2005,** *93*, 1451-1458.

60.      Suo, Z.; Ma, E. Y.; Gleskova, H.; Wagner, S. Mechanics of Rollable and Foldable Film-on-Foil Electronics. *Appl. Phys. Lett.* **1999,** *74*, 1177-1179.

61.     Khan, A.; Lee, S.; Jang, T.; Xiong, Z.; Zhang, C.; Tang, J.; Guo, L. J.; Li, W.-D. High-Performance Flexible Transparent Electrode with an Embedded Metal Mesh Fabricated by Cost-Effective Solution Process. *Small* **2016,** *12*, 3021-3030.

62.     Kiruthika, S.; Gupta, R.; Kulkarni, G. U. Large Area Defrosting Windows Based on Electrothermal Heating of Highly Conducting and Transmitting Ag Wire Mesh. *RSC Advances* **2014,** *4*, 49745-49751.

63.     Kang, J.; Jang, Y.; Kim, Y.; Cho, S. H.; Suhr, J.; Hong, B. H.; Choi, J. B.; Byun, D. An Ag-Grid/Graphene Hybrid Structure for Large-Scale, Transparent, Flexible Heaters. *Nanoscale* **2015,** *7*, 6567-6573.

64.     Kim, T.; Kim, Y. W.; Lee, H. S.; Kim, H.; Yang, W. S.; Suh, K. S. Uniformly Interconnected Silver-Nanowire Networks for Transparent Film Heaters. *Adv. Funct. Mater.* **2013,** *23*, 1250-1255.

65.     Smith, R. T.; Chern, M. J.; Hegg, R. G. Automotive Head-up Display with High Brightness in Daytime and High Contrast in Nighttime. Patent US5053755 A, Oct 1, 1991.

66.     Flir One for Android Specification. http://www.flir.com/flirone/android/ (accessed 21 July, 2016).

67.     Yoon, Y. H.; Song, J. W.; Kim, D.; Kim, J.; Park, J. K.; Oh, S. K.; Han, C. S. Transparent Film Heater Using Single-Walled Carbon Nanotubes. *Adv. Mater.* **2007,** *19*, 4284-4287.

68.     Bae, J. J.; Lim, S. C.; Han, G. H.; Jo, Y. W.; Doung, D. L.; Kim, E. S.; Chae, S. J.; Huy, T. Q.; Van Luan, N.; Lee, Y. H. Heat Dissipation of Transparent Graphene Defoggers. *Adv. Funct. Mater.* **2012,** *22*, 4819-4826.

69.     Kang, T. J.; Kim, T.; Seo, S. M.; Park, Y. J.; Kim, Y. H. Thickness-Dependent Thermal Resistance of a Transparent Glass Heater with a Single-Walled Carbon Nanotube Coating. *Carbon* **2011,** *49*, 1087-1093.

70.     Sui, D.; Huang, Y.; Huang, L.; Liang, J.; Ma, Y.; Chen, Y. Flexible and Transparent Electrothermal Film Heaters Based on Graphene Materials. *Small* **2011,** *7*, 3186-3192.

**Table of contents graphic**

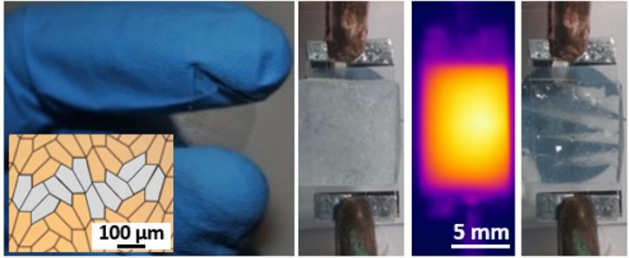